\documentclass[slac_one]{revtex4}
\usepackage{graphicx}
\usepackage{fancyhdr}
\usepackage{atlasphysics}

\begin{document}

\title{Observation of a new $\chib$ state at ATLAS and a new $\Xi_{b}$ baryon at CMS} 

%

\author{Andrew S. Chisholm, on behalf of the ATLAS and CMS Collaborations}
\affiliation{School of Physics and Astronomy, University of Birmingham, Birmingham, United Kingdom}

\begin{abstract}

Two recent observations of new $b$ hadrons in $pp$ collisions at $\rts = 7 \TeV$ at the Large Hadron Collider (LHC) are presented. The ATLAS collaboration has observed a new state in radiative transitions to ${\Ups}(1S)$ and ${\Ups}(2S)$ and interprets this as the first observation of the $\chib(3P)$ states. The CMS collaboration has observed a new $b$ baryon decaying to $\Xi_{b}^{-}\pi^{+}$ (plus charge conjugates). This is interpreted as a neutral $J^{P} = 3/2^{+}$ $\Xi_{b}^{*}$ baryon.

\end{abstract}

\maketitle

\thispagestyle{fancy}


\section{INTRODUCTION} 
\noindent The high cross section for heavy flavour production at the Large Hadron Collider (LHC) offers a fantastic opportunity to study the known heavy flavour hadrons and to search for the many ``missing'' states predicted in the quark model. In particular, hadrons containing a $b$ quark are expected to be copiously produced at the LHC. Further to this, many heavy hadron states (bottom baryons in particular) are only accessible at hadron colliders, and with the end of data taking at the Tevatron, the LHC is fast becoming the one of the best facilities to study heavy hadron spectroscopy. The LHC has already delivered its first two discoveries of new $b$ hadrons within the first three years of running. The ATLAS collaboration has observed a new state in radiative transitions to ${\Ups}(1S)$ and ${\Ups}(2S)$ and interprets this as the first observation of the $\chib(3P)$ states~\cite{ATLAS_chib}. The CMS collaboration observes a new $b$ baryon decaying to $\Xi_{b}^{-}\pi^{+}$ (plus charge conjugates), interpreted as a neutral $\Xi_{b}^{*}$ baryon~\cite{CMS_Xi}. 

\section{OBSERVATION OF A NEW $\chib$ STATE AT ATLAS}
\noindent The $\chib$ states represent the $S=1$ (parallel quark spins) $P$-wave states of the bottomonium ($\bbbar$) system. The $\chib$ comprise a triplet of states,  $\chibj0$, $\chibj1$, $\chibj2$, with quantum numbers $J^{PC}=0^{++}, 1^{++}, 2^{++}$. The three states are characterised by a small hyperfine mass splitting of ${\cal{O}}(10 \MeV)$. The branching fractions for the radiative decays $\chib \to {\Ups} \, \gamma$ are large, ${\cal{O}}(10 \%)$. The $\chib(1P)$ and $\chib(2P)$ triplets (with spin-weighted masses of around $9.90$ and $10.26$ GeV respectively) have been studied in detail at $\ee$ colliders, providing precise measurements of the hyperfine mass structure and radiative branching fractions \cite{chib_ee1, chib_ee2, chib_ee3}. Additionally, a third triplet of states, the $\chib(3P)$, is expected below the open beauty threshold at a mass around $10.525\GeV$~\cite{predictions1, predictions2, predictions3}. 

Given the complex hadronic environment of the LHC, the radiative decays $\chib \to {\Ups} \, \gamma$ with ${\Ups}\to\mumu$ represent a very clean experimental channel to study the $\chib$ states, where the presence of two muons offers a clear signature to trigger upon. The aim of the ATLAS analysis is to reconstruct the radiative decays $\chib\to{\Ups}(1S)\gamma$ and $\chib\to{\Ups}(2S)\gamma$ with two independent analyses that reconstruct the photon with either a direct calorimetric measurement or through the reconstruction of $\ee$ conversions in the ATLAS tracker. The two photon reconstruction methods have their own advantages and disadvantages. In particular, converted photons offer better invariant mass resolution compared to photons reconstructed by the calorimeter but at the expense of a much lower reconstruction efficiency.  The ATLAS collaboration has recently reported the observation of a new structure decaying to ${\Ups}(1S)\gamma$ and ${\Ups}(2S)\gamma$ consistent with the $\chib(3P)$ system~\cite{ATLAS_chib}. The following sections summarise the ATLAS analysis and results, a detailed description of the ATLAS detector can be found in \cite{ATLAS_Paper}. 

\subsection{Data Sample and Selection of ${\Ups} \to \mumu$ decays}
\label{sec_ATLAS_Upsi}
\noindent The ATLAS analysis uses a data sample, recorded by the ATLAS experiment during the 2011 LHC proton-proton collision run at a centre of mass energy $\rts = 7 \TeV$, representing an integrated luminosity of $4.4\,\ifb$. The data sample was collected by a set of triggers designed to select events containing di-muon candidates or single high transverse momentum muons.

Muon candidates are reconstructed from stand alone tracks reconstructed in the ATLAS muon spectrometer combined with tracks reconstructed in the ATLAS Inner Detector (ID). Two oppositely charged muon candidates, both with transverse momentum $p_{T} > 4 \GeV$ and pseudorapidity $|\eta| < 2.3$, are fitted to a common vertex to form di-muon candidates. The di-muon vertex fit quality is required to satisfy $\chi^{2}/d.o.f < 20$. Finally, di-muon candidates are required to have transverse momentum $p_{T} > 12 \GeV$ and rapidity $|y|< 2.0$. The invariant mass distribution of di-muon candidates  is shown in Figure~\ref{Upsilon}. ${\Ups}(1S)$ and ${\Ups}(2S)$ candidates are selected from di-muon candidates with an invariant mass within the range $9.25 < m({\mumu}) < 9.65 \GeV$ and $9.80 < m({\mumu}) < 10.10 \GeV$ respectively. 

\begin{figure*}[t]
\centering
\includegraphics[width=135mm]{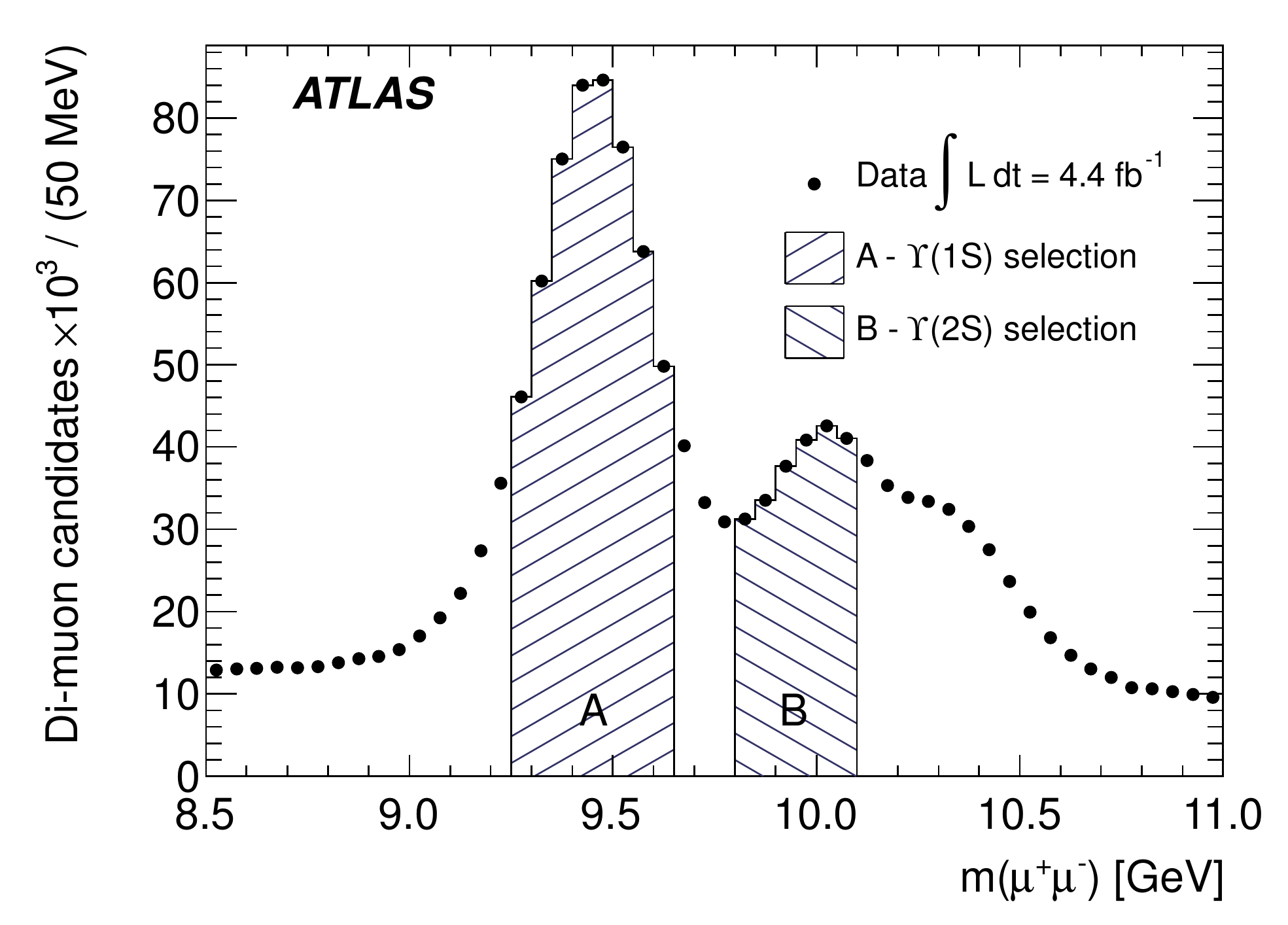}
\caption{The invariant mass distribution of di-muon candidates selected as described in Section~\ref{sec_ATLAS_Upsi}. The shaded areas A and B represent the invariant mass selections for ${\Ups}(1S)$ and ${\Ups}(2S)$ candidates respectively~\cite{ATLAS_chib}.} \label{Upsilon}
\end{figure*}

\subsection{Converted Photon Selection}
\noindent Candidate photon conversions are reconstructed from two oppositely charged tracks reconstructed in the ATLAS inner detector (ID) that intersect at a common vertex. The two tracks are fitted to a common conversion vertex where the fit is required to converge with a $\chi^{2}$ probability greater than 0.01. Each electron track is required to have transverse momentum $p_{T} > 500 \MeV$ and pseudorapidity $|\eta| < 2.3$ and to be reconstructed from at least 4 hits in the silicon layers of the ID. Conversion candidates reconstructed from tracks associated to the the di-muon candidate are rejected. To reduce background contamination from Dalitz decays and fake conversions, conversion vertices are required to be reconstructed with a radial distance from the beam axis of greater than $40\,\mathrm{mm}$. Converted photons not associated with the di-muon vertex are rejected by demanding that the impact parameter of the converted photon candidate with respect to the di-muon vertex be less than $2\,\mathrm{mm}$. 

\subsection{Unconverted Photon Selection}
\noindent Unconverted photons are reconstructed from energy deposits in the ATLAS electromagnetic calorimeter that are not matched to any ID track. Unconverted photon candidates are required to have a transverse energy greater than $2.5\GeV$ and pseudorapidity $|\eta| < 2.37$. Photon candidates are also required to satisfy the ``loose'' photon identification selection described in \cite{ATLAS_photon} to reject backgrounds from $\pi^{0}\to\gamma\gamma$ decays and narrow jets. Unconverted photons reconstructed within the transition region ($1.37 < |\eta| < 1.52$) between the barrel and end cap calorimeters are not selected. To improve the momentum resolution of unconverted photons, the polar angle of the unconverted photon is corrected to point back to the di-muon vertex, exploiting the longitudinal segmentation of the ATLAS electromagnetic calorimeter. This procedure also allows photons that are not compatible with having originated from the di-muon vertex to be rejected through a loose cut on the fit quality of $\chi^{2}/d.o.f < 200$. This procedure is described in detail in \cite{ATLAS_pointing}.

\subsection{Selection of $\chib$ Candidates}
\noindent Reconstructed ${\Ups} \to \mumu$ candidates are associated with reconstructed photons to form $\chib$ candidates.  To minimise the effects of the experimental di-muon mass resolution, the invariant mass difference $\Delta m = m(\mumu\gamma)-m(\mumu)$ is calculated. The $\Delta m$ distributions for ${\Ups}(1S)\gamma$ and ${\Ups}(2S)\gamma$ candidates can be shown on the same mass scale through the definition of the $\Delta m + m_{{\Ups}(kS)}$ distribution (the $m_{{\Ups}(k=1,2S)}$ represent the current world average masses of the ${\Ups}(1S)$ and ${\Ups}(2S)$ states)~\cite{PDG}. The $\Delta m + m_{{\Ups}(1,2S)}$ distributions for $\chib$ candidates reconstructed from unconverted and converted photons are shown in Figures \ref{chib_calo} and  \ref{chib_conv} respectively. A final selection requirement on the transverse momentum of the di-muon system of $p_{T} > 20 \GeV$ is imposed for unconverted photon candidates to maximise the signal significance of the $\chib(1P)$ and $\chib(2P)$ peaks.

\begin{figure*}[t]
\centering
\includegraphics[width=135mm]{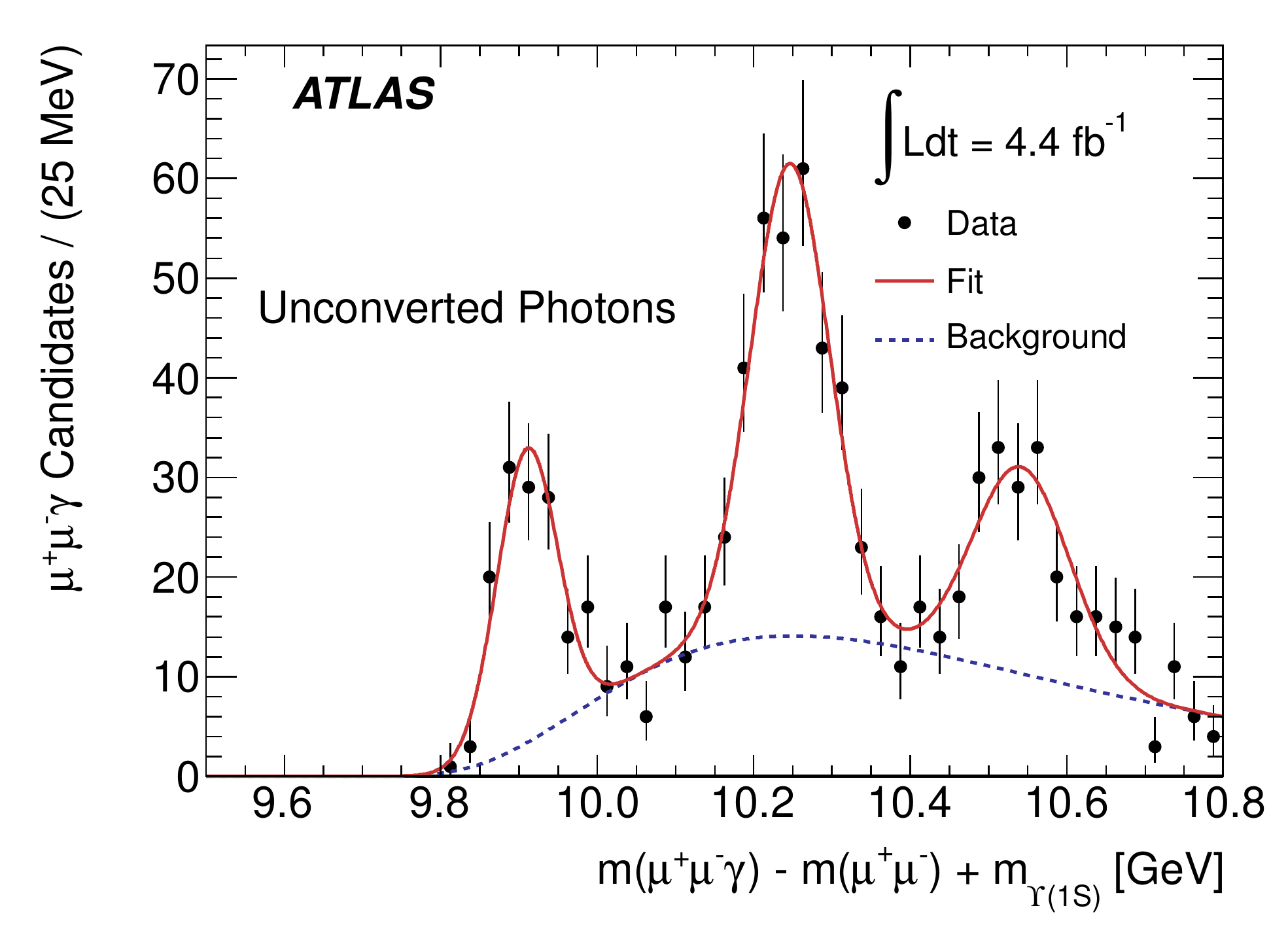}
\caption{The $\Delta m + m_{{\Ups}(1S)}$ distribution for $\chib\to{\Ups}(1S)\gamma$ candidates reconstructed from unconverted photons~\cite{ATLAS_chib}.} \label{chib_calo}
\end{figure*}

\begin{figure*}[t]
\centering
\includegraphics[width=135mm]{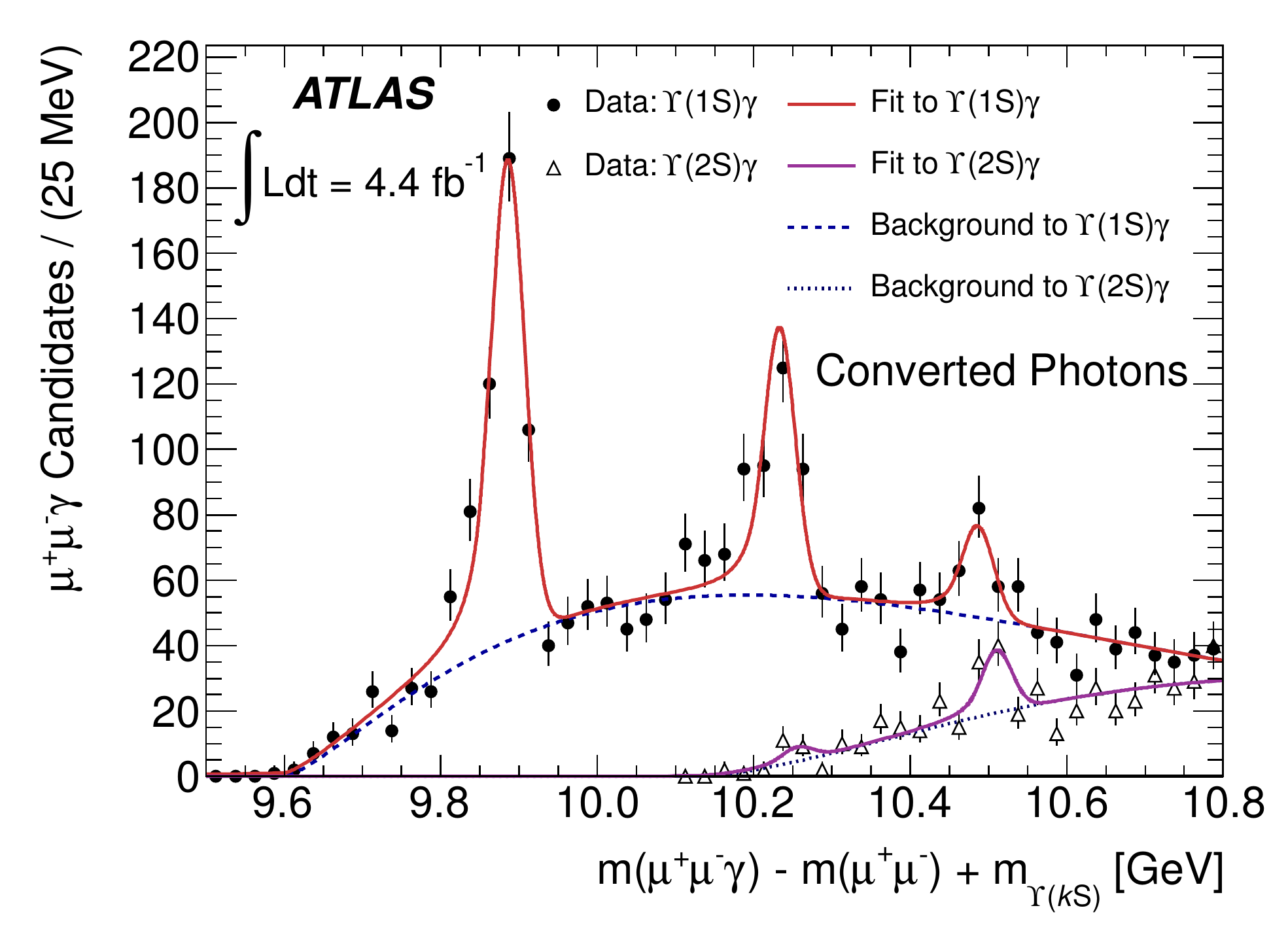}
\caption{The $\Delta m + m_{{\Ups}(1S)}$ (filled points) and $\Delta m + m_{{\Ups}(2S)}$ (open triangles) distributions  for $\chib\to{\Ups}(1S,2S)\gamma$ candidates reconstructed from converted photons. The data points have not been corrected for energy losses due to bremsstrahlung~\cite{ATLAS_chib}.} \label{chib_conv}
\end{figure*}

The mass distributions for ${\Ups}(1S)\gamma$ candidates reconstructed from both unconverted and converted photons shown in Figures \ref{chib_calo} and  \ref{chib_conv} exhibit clear peaks at approximately $9.9 \GeV$ and $10.2 \GeV$ consistent with $\chib(1P)\to{\Ups}(1S)\gamma$ and $\chib(2P)\to{\Ups}(1S)\gamma$ decays. In addition to these peaks, a third structure is also observed at a mass of approximately $10.5 \GeV$. This additional structure is also observed in the mass distribution for ${\Ups}(2S)\gamma$ candidates. The mass and decay modes of these additional structures is consistent with the expectations for the $\chib(3P)$ states decaying in the modes $\chib(3P)\to{\Ups}(1S)\gamma$ and $\chib(3P)\to{\Ups}(2S)\gamma$. The higher transverse momentum threshold for the reconstruction of unconverted photons prohibits the reconstruction of soft photons from $\chib(2P,3P)\to{\Ups}(2S)\gamma$ decays.

\subsection{Fit Description and Results}
\noindent Unbinned maximum likelihood fits are performed to the $\Delta m + m_{{\Ups}(1,2S)}$ distributions for both the converted and unconverted $\chib$ candidates to measure the mass of the new structure under its interpretation as the $\chib(3P)$ states, both fits are described in detail in~\cite{ATLAS_chib}. 

The $\Delta m + m_{{\Ups}(1S)}$ distribution for unconverted photon candidates is described by three Gaussian probability density functions (PDFs), each with independent mean value, width and normalisation parameters. The background distribution is described by the smooth function, $\exp{\left(A\Delta m + B {\Delta m}^{-2}\right)}$ with two free parameters, $A$ and $B$. The mass barycenter of the $\chib(3P)$ signal is measured to be $\overline{m}_{3} = 10.541 \pm 0.011 \stat \pm 0.030 \syst \GeV$ from the fit to unconverted photon candidates alone~\cite{ATLAS_chib}. The systematic uncertainty on the unconverted photon mass measurement is dominated by the modelling of the background distribution and the uncertainty associated with the unconverted photon energy scale.

Both the $\Delta m + m_{{\Ups}(1S)}$ and $\Delta m + m_{{\Ups}(2S)}$ distributions for converted photon candidates are fitted together in a simultaneous fit. Each $\chib(nP)$ peak is described by a pair of Crystal Ball (CB) functions. This is motivated by the fact that the converted photon mass resolution is comparable to the hyperfine splitting between the $J=1,2$ states. The background distribution for both the $\Delta m + m_{{\Ups}(1S)}$ and $\Delta m + m_{{\Ups}(2S)}$ distributions is described by the function $(\Delta m -  q_{0})^{\alpha}\cdot \exp\left \{ (\Delta m -  q_{0})\cdot \beta \right \}$ where $q_{0}$, $\alpha$ and $\beta$ are all free parameters (an independent set of parameters for both the $\Delta m + m_{{\Ups}(1,2S)}$ distributions). The mass barycenter of the $\chib(3P)$ signal, determined from converted photon candidates alone, is measured to be $\overline{m}_{3} = 10.530 \pm 0.005\stat \pm 0.009\syst \GeV$~\cite{ATLAS_chib}. The systematic uncertainty associated with the mass measurement is dominated by the various assumptions made in the simultaneous fit.

The mass measurements using both converted and unconverted photons are compatible with each other and with the theoretical expectations for the $\chib(3P)$ system. The mass measurement from the converted photon analysis with the lower statistical and systematic uncertainty is chosen to represent the final measurement of the mass barycenter of the $\chib(3P)$ system.

The significance of the $\chib(3P)$ signal is assessed from the logarithmic likelihood ratio $\log{\left(L_{max}/L_{0}\right)}$, where $L_{max}$ and $L_{0}$ represent likelihood values calculated from fits with and without a $\chib(3P)$ signal included respectively. The significance is re-assessed for each set of systematic variations and is consistently found to be in excess of $6$ standard deviations for both the unconverted and converted photon analyses independently.

The observation of a new structure in the ${\Ups}(1S)\gamma$ spectrum was recently confirmed by the  $D\O$ collaboration~\cite{D0}.  $D\O$ measure the mass of the structure to be $\overline{m}_{3} = 10.551 \pm 0.014\stat \pm 0.017\syst \GeV$, consistent with the ATLAS measurement.

\subsection{Conclusion}
\noindent The ATLAS collaboration has observed the known $\chib(1P)$ and $\chib(2P)$ states in radiative transitions to ${\Ups}(1S)\gamma$ in proton-proton collisions at the LHC. In addition to this, ATLAS observe a new structure in radiative decays to ${\Ups}(1S)\gamma$ and ${\Ups}(2S)\gamma$ consistent with theoretical expectations for the $\chib(3P)$ system. The mass barycenter of this structure, under the interpretation as the $\chib(3P)$ system, is measured to be $\overline{m}_{3} = 10.530 \pm 0.005\stat \pm 0.009\syst \GeV$~\cite{ATLAS_chib}. Further measurements by ATLAS and the other LHC experiments will hopefully shed more light on this new state in the near future.


\section{OBSERVATION OF A NEW $\Xi_{b}$ BARYON AT CMS}

\noindent Until recently, the $\Xi_{b}$ states represented the only experimentally observed baryons to contain one strange and one bottom valence quark within the quark model of baryons. The first direct observation of the $\Xi_{b}$, of which there exists both neutral $usb$ and negatively charged $dsb$ varieties, came from the Tevatron experiments ~\cite{Tev1,Tev2,Tev3}, although indirect evidence for the $\Xi_{b}^{-}$ was seen at LEP~\cite{LEP1,LEP2}. In addition to the the $J^{P} = 1/2^{+}$ $\Xi_{b}$ ground states, the quark model predicts the $J^{P} = 1/2^{+}$ $\Xi_{b}^{\prime}$, $J^{P} = 3/2^{+}$ $\Xi_{b}^{*}$ and two further states with $L=1$ orbital angular momentum between the $b$ quark and the light di-quark system~\cite{XiTheory1,XiTheory2,XiTheory3,XiTheory4,XiTheory5}. The strong decay $\Xi_{b}^{\prime}\to \Xi_{b}\pi$ is expected to be kinematically forbidden due to an expected $\Xi_{b}^{\prime}-\Xi_{b}$ mass difference below the pion mass. However, the $\Xi_{b}^{*}\to \Xi_{b}\pi$ decay is expected to be kinematically allowed, in analogy with the better studied $\Xi_{c}$ baryon system.

The CMS analysis represents a search for $\Xi_{b}^{*0}$ baryons in $\Xi_{b}^{-}\pi^{+}$ (plus charge conjugate) decays with $\Xi_{b}^{-}\to\Jpsi\,\Xi^{-}$, $\Jmumu$, $\Xi^{-}\to \Lambda^{0}\,\pi^{-}$ and $\Lambda^{0}\to p\pi^{-}$ (plus charge conjugates). For brevity, the reconstruction of the charge conjugate decay modes will be implied throughout this summary. The following sections summarise the CMS analysis and results presented in \cite{CMS_Xi}. A detailed description of the CMS detector can be found in~\cite{CMS_Paper}.

\subsection{Data Sample and Event Selection}

\noindent The CMS analysis is based on a sample of $pp$ collision data representing an integrated luminosity of $5.3\,\ifb$ collected during the 2011 LHC run at a centre of mass energy $\rts = 7\TeV$. The data used by the CMS analysis are collected by specialised triggers designed to record events containing two oppositely charged muons that are compatible with having been produced in the decay of a $\Jpsi$. Separate triggers are used to select events containing $\Jmumu$ candidates that are promptly produced and those that are displaced from the primary vertex.

\subsection{Selection of $\Xi_{b}^{-}$ candidates}

\noindent The foundation of the analysis is the reconstruction of a pure sample of $\Xi_{b}^{-}\to\Jpsi\,\Xi^{-}$ decays (with $\Xi^{-}\to \Lambda^{0}\,\pi^{-}$ and $\Lambda^{0}\to p\pi^{-}$). Candidate $\Jpsi$ are formed from pairs of oppositely charged muons reconstructed from tracks in the silicon tracker, matched to independent tracks in the muon detectors. The two muons are required to pass the trigger selection and the di-muon candidate must have an invariant mass within $150\MeV$ of the world average $\Jpsi$ mass~\cite{PDG}.

The reconstruction of $\Xi^{-}$ decays begins with the identification of candidate $\Lambda^{0}\to p\pi$ decays from pairs of oppositely charged tracks. The track with the higher momentum is taken to be the proton. The two tracks are required to have been reconstructed from at least 6 six hits in the silicon tracker and to have a track fit $\chi^{2}/d.o.f < 5$. The two tracks are fitted to a common decay vertex and the fit result is required to satisfy  $\chi^{2}/d.o.f < 7$. The decay vertex is required to be displaced from the beam line by more than ten times the uncertainty on the displacement. To remove possible contamination from $K^{0}_{s}$ decays, the invariant mass of the candidate (with both tracks assigned the pion mass) is rejected if it lies within $20\MeV$ of the $K^{0}_{s}$ mass. $\Xi^{-}$ candidates are then reconstructed through the combination of a candidate $\Lambda^{0}$ with a track (denoted $\pi_{\Xi}$) with the same charge as the pion from the $\Lambda^{0}\to p\pi$ candidate (denoted $\pi_{\Lambda}$). The three tracks are then subjected to a kinematic vertex fit where the invariant mass of the two tracks from the $\Lambda^{0}$ candidate are constrained to the world average $\Lambda^{0}$ mass. To reject backgrounds from mis-reconstructed $\Omega^{-}\to\Lambda^{0}\,K^{-}$ decays, the $\Xi^{-}$ candidate is rejected if its invariant mass, with the $\pi_{\Xi}$ track given the charged kaon mass, lies within $20\MeV$ of the $\Omega^{-}$ mass.

Finally, $\Xi_{b}^{-}$ candidates are formed by combining candidate $\Xi^{-}$ and $\Jpsi$ decays in a kinematic vertex fit where the masses of the $\Xi^{-}$and $\Jpsi$ candidates are constrained to the world average values.

To select a high yield sample of $\Xi_{b}^{-}$ candidates with high purity, the $\Xi_{b}^{-}$ selection (characterised by the values of thirty selection variables) is optimised by an iterative algorithm that maximises both the signal yield and significance. The thirty optimised variables include the transverse momentum threshold of the $\Jpsi$, $p$, $\pi_{\Lambda}$, $\pi_{\Xi}$, $\Xi^{-}$, $\Xi_{b}^{-}$ and muon candidates. The values of the latter two thresholds ($\Xi_{b}^{-}$ and both muons) can take different values depending on whether the candidates were reconstructed in the barrel or endcap regions on the detector. The requirements on the difference between the reconstructed invariant masses and the world average values for the $\Lambda^{0}$, $\Xi^{-}$ and $\Jpsi$ candidates are also optimised. Other important optimised variables include the impact parameter significance of the $p$, $\pi_{\Lambda}$ and $\pi_{\Xi}$ tracks and transverse decay length (and its significance) of the $\Lambda^{0}$, $\Xi^{-}$ and $\Xi_{b}^{-}$ decay vertices. The details of the optimisation procedure are described in \cite{CMS_Xi}.

Figure~\ref{XiB} shows the invariant mass of selected $\Xi_{b}^{-}$ candidates. The $m(\Jpsi\,\Xi^{-})$ distribution exhibits a clear peak at approximately $5.8\GeV$, consistent with a $\Xi_{b}^{-}$ signal. The $m(\Jpsi\,\Xi^{-})$ distribution is fit with a Gaussian PDF to describe the signal and a second order polynomial to describe the background. The fit results in a yield of $108\pm14$ $\Xi_{b}^{-}$ candidates and a fitted $\Xi_{b}^{-}$ mass of $5795.0\pm3.1\stat \MeV$, a value which agrees well with the current world average~\cite{PDG}.

\begin{figure*}[t]
\centering
\includegraphics[width=135mm]{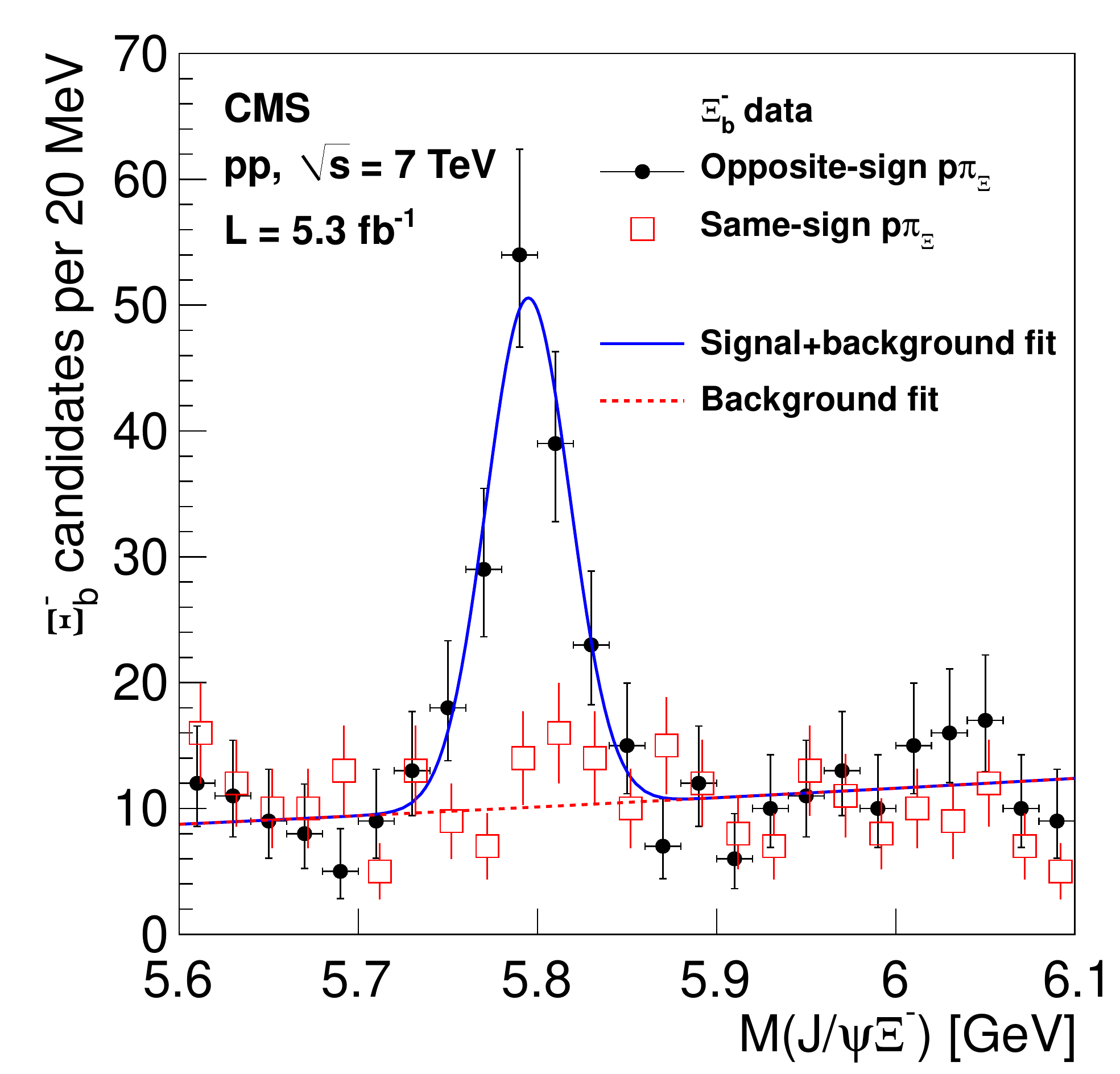}
\caption{The $\Jpsi\,\Xi^{-}$ invariant mass distribution for $\Xi_{b}^{-}$ candidates (filled points). The invariant mass distribution of $\Jpsi\,\Xi^{-}$ pairs where the proton and $\pi_{\Xi}$ have the same charge is shown also shown (open squares)~\cite{CMS_Xi}. } \label{XiB}
\end{figure*}

\subsection{Search for $\Xi_{b}^{*0}$ baryons}

\noindent $\Xi_{b}^{-}$ candidates with a mass within $2.5\sigma$ of the fitted $\Xi_{b}^{-}$ mass are associated with a track, given the pion mass, with a charge opposite to that of the $\pi_{\Xi}$. The tracks are required to be compatible with having been produced at the primary vertex and to have a transverse momentum greater than $0.25\GeV$. The events in the CMS data sample contain on average eight primary vertices. The primary vertex reconstructed closest to the $\Xi_{b}^{-}$ line of flight is assumed to be associated with the production of the $\Xi_{b}^{-}$. 

A potential $\Xi_{b}^{*0}$ signal is expected to appear as a peak in the $Q = m( \Jpsi\,\Xi^{-}\pi^{+}) - m(\Jpsi\,\Xi^{-}) - m(\pi)$ distribution. In order to search for such peaks, the background contribution to the $Q$ distribution must first be reliably estimated. This is done through the preparation of a background sample of $\Xi_{b}^{-}$ candidates associated with prompt pion tracks of the same charge as the $\Xi_{b}^{-}$. The $Q$ distribution for this background sample is shown in Figure~\ref{XiB_SameSign}. The momentum distributions of the $\Xi_{b}^{-}$ and pions ($p(\Xi_{b})$ and $p(\pi)$) and the distribution of angle between them ($\alpha$) from the (same sign) background sample are used to randomly generate an uncorrelated set of values for $p(\Xi_{b})$, $p(\pi)$ and $\alpha$. The uncorrelated set of values is used to calculate a value for $Q$, this is repeated $10^8$ times to give a $Q$ distribution which is expected to predict the shape of the combinatorial background. This distribution is then fitted with the function $Q^{c_{1}}\left( e^{-c_{2}Q} + e^{-c_{3}Q} + e^{-c_{4}Q} \right)$ where the $c_{i}$ are all free parameters. The fit result is shown by the red dashed line in Figure~\ref{XiB_SameSign}. 

\begin{figure*}[t]
\centering
\includegraphics[width=135mm]{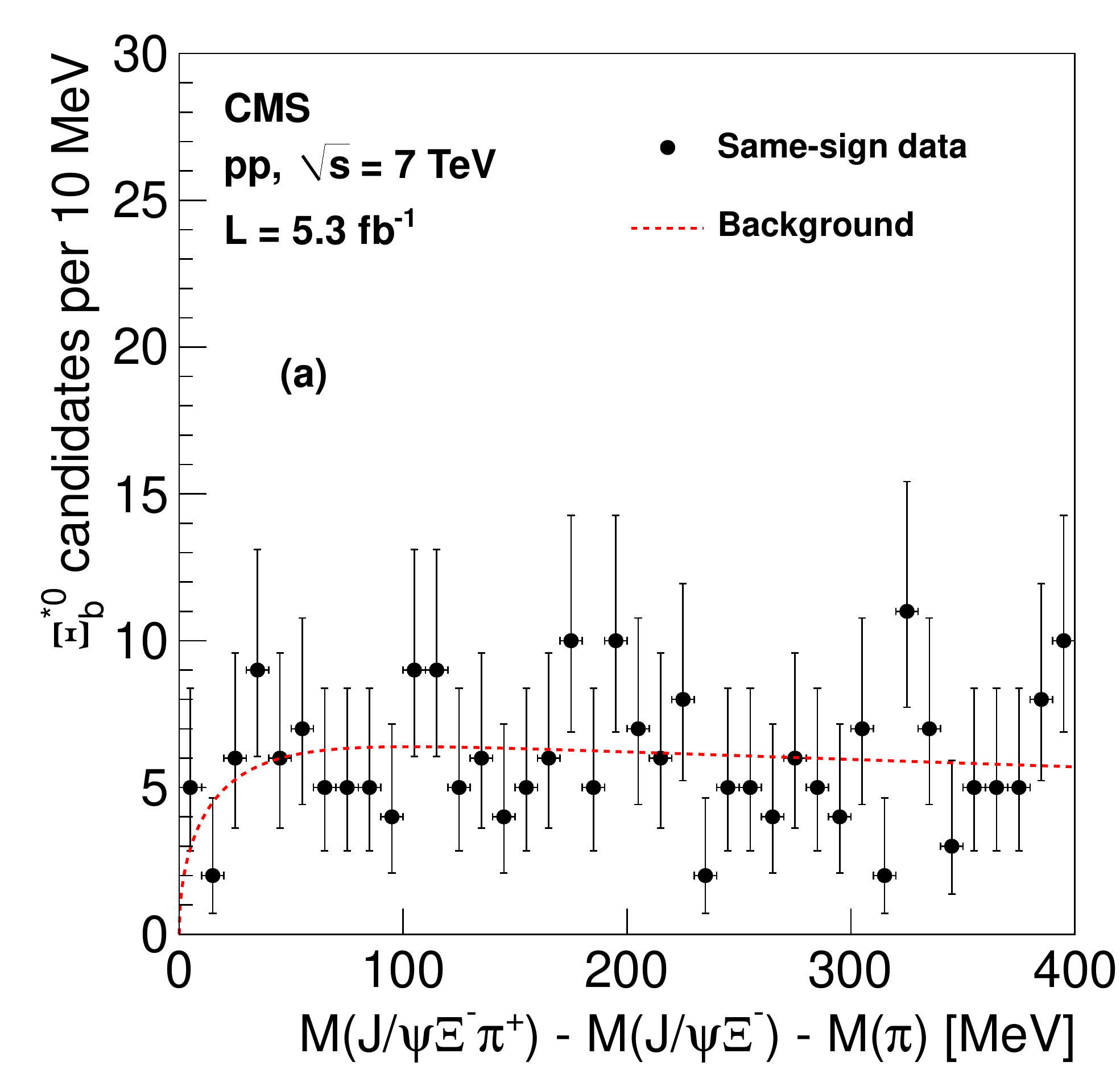}
\caption{The $Q$ distribution for $\Xi_{b}^{-}$ candidates associated with prompt pions of the same charge~\cite{CMS_Xi}.} \label{XiB_SameSign}
\end{figure*}

\begin{figure*}[t]
\centering
\includegraphics[width=135mm]{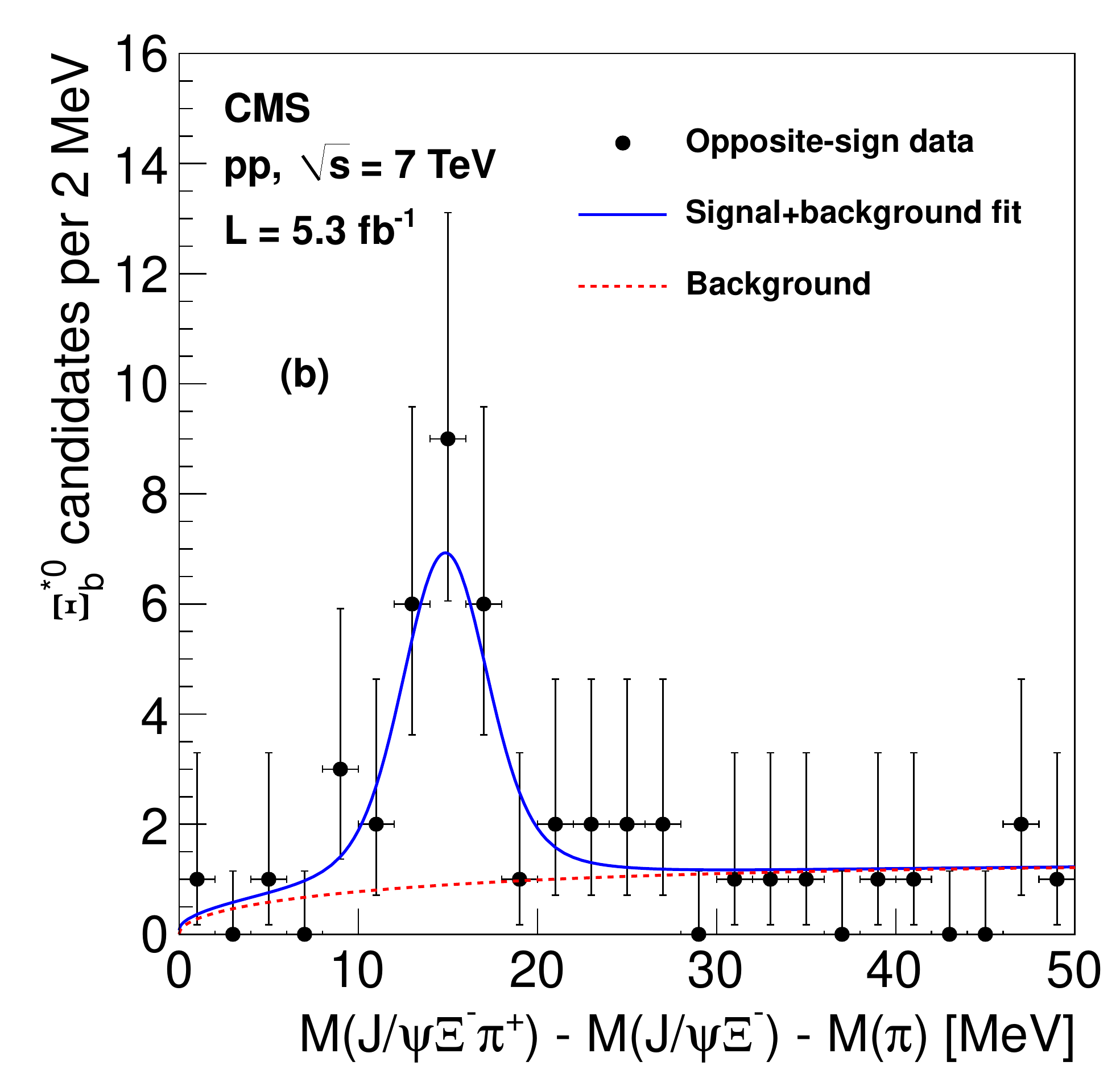}
\caption{The $Q$ distribution for $\Xi_{b}^{-}\pi^{+}$ candidates where the $\Xi_{b}^{-}$ and prompt pion candidates have opposite charges~\cite{CMS_Xi}.} \label{XiBStar}
\end{figure*}

The $Q$ value distribution for opposite sign $\Xi_{b}^{-}\pi^{+}$ combinations is shown in Figure~\ref{XiBStar}. The $Q$ value distribution in Figure~\ref{XiBStar} exhibits a clear excess above the background expectation in the region $12 < Q < 18 \MeV$. An unbinned maximum likelihood fit is performed to the $Q$ value distribution in Figure~\ref{XiBStar} where the excess is modelled by a Breit-Wigner distribution convolved with a Gaussian resolution function (with a width fixed to $1.91\pm0.11\MeV$, a value derived from Monte Carlo (MC) simulation). The background model discussed earlier is used to describe the combinatorial background where the $c_{i}$ parameters are free to vary within their total uncertainties. The fitted mean value of the signal is $14.84\pm0.74\stat\MeV$ with a Breit-Wigner width of $2.1\pm1.7\stat\MeV$~\cite{CMS_Xi}.

The significance of the signal is assessed though the calculation of $\sqrt{\ln{\left(L_{s+b}/L_{b}\right)}}$ where $L_{s+b}$ is the likelihood value of the nominal (signal and background) fit and $L_{b}$ is the likelihood value for a fit performed with a background only model. The statistical significance calculated in the way is $6.9$ standard deviations. The significance is also evaluated though a number of other approaches (including a treatment of the ``look elsewhere effect''), all of which result in a significance in excess of $5$ standard deviations.

Various cross checks are performed using MC simulation samples of $B^{+}$, $B^{0}$, $B_{s}$ and $\Lambda_{b}$ decays. The analysis procedure performed on these samples does not exhibit any excess in the $Q$ distribution that might arise from the mis-reconstruction of known $b$ hadrons.

The systematic uncertainty on the measured $Q$ value of the signal receives contributions from the observation of a small upwards shift in the $Q$ distribution in MC simulation and uncertainties associated with the background model. Together these effects constitute a systematic uncertainty of $0.28\MeV$.

With the charged pion and $\Xi_{b}^{-}$ masses taken from the PDG the mass of the new baryon is measured to be $5945.0 \pm 0.7\stat \pm 0.3\syst \pm 2.7~(\mathrm{PDG}) \MeV$~\cite{CMS_Xi}.
\subsection{Conclusion}

\noindent The CMS collaboration has observed a new $\Xi_{b}$ baryon, decaying to $\Xi_{b}^{-}\pi^{+}$ (plus charge conjugate) with a statistical significance in excess of $5$ standard deviations. The mass of the new baryon is measured to be $5945.0 \pm 0.7\stat \pm 0.3\syst \pm 2.7~(\mathrm{PDG}) \MeV$~\cite{CMS_Xi}. The measured mass and decay mode are consistent with theoretical expectations for the $\Xi_{b}^{*0}$ state, predicted to have quantum numbers $J^{P} = 3/2^{+}$.

\clearpage

\begin{acknowledgments}
\noindent The author is grateful for the financial support provided by the Science and Technology Facilities Council (STFC) in the UK.
\end{acknowledgments}


\end{document}